\documentclass[12pt]{article} \input epsf.tex
\usepackage{epsfig,graphicx,psfrag}
\usepackage{axodraw}
\usepackage{amsmath,bbm,amsfonts,amssymb}
\usepackage{latexsym}\usepackage{stmaryrd}

\newcommand{\be}{\begin{equation}}
\newcommand{\ee}{\end{equation}}
\newcommand{\bea}{\begin{eqnarray}}
\newcommand{\eea}{\end{eqnarray}}
\newcommand{\bear}{\begin{eqnarray}}
\newcommand{\eear}{\end{eqnarray}}
\newcommand{\ba}{\begin{array}}
\newcommand{\ea}{\end{array}}

\newcommand{\beq}{\begin{equation}}
\newcommand{\eeq}{\end{equation}}
\newcommand{\beqs}{\begin{eqnarray}}
\newcommand{\eeqs}{\end{eqnarray}}

\begin{document}

\baselineskip=18pt \pagestyle{plain} \setcounter{page}{1}

\vspace*{-0.8cm}

\noindent \makebox[11.6cm][l]{\small \hspace*{-.2cm} December 17, 2009}
{\small Fermilab-Pub-09-630-T}
\\ [-2mm]

\begin{center}
{\Large \bf Semileptonic decays of the standard Higgs boson \\
 [9mm]

{\normalsize \bf Bogdan A. Dobrescu and Joseph D. Lykken 
\\ [4mm]
{\small {\it
Theoretical Physics Department, Fermilab, Batavia, IL 60510, USA }}\\
}}
\end{center}

\vspace*{0.1cm}

\begin{abstract}
The Higgs boson decay into a pair of real or virtual $W$ bosons, with one of them 
decaying leptonically, is predicted within the Standard Model to have the largest branching fraction
of all Higgs decays that involve an isolated electron or muon, for $M_h > 120$ GeV.
We compute analytically the fully-differential width for this $h^0\! 
\to \! \ell\nu jj$ decay at tree level, and then explore some multi-dimensional cuts 
that preserve the region of large signal. 
Future searches for semileptonic decays at the Tevatron and LHC, 
employing fully-differential information as outlined here,
may be essential for ruling out or in the Higgs boson and for characterizing a Higgs signal.
\end{abstract}

\smallskip

\section{Introduction} \setcounter{equation}{0}

The most pressing challenge in particle physics is to understand the origin of 
electroweak symmetry breaking. The proximate experimental question is whether a Higgs boson exists;
the Standard Model predicts the existence of a Higgs boson whose properties are entirely 
determined by its mass $M_h$ \cite{Djouadi:2005gi}. 
The LEP bound $M_h \ge 114.4$ GeV \cite{Amsler:2008zzb}, set a decade ago, has 
been extended at the Tevatron, where the $163 - 166$ GeV range for $M_h$
has recently been ruled out at the 95\% confidence level \cite{Collaboration:2009je}.
This tour de force relied entirely on sensitivity to the leptonic decay 
chain analyzed in Refs.~\cite{Han:1998ma}-\cite{Carena:2000yx}:
$h^0\! \to W^+W^-  \!\!\to \ell^+\nu\ell^-\bar{\nu}$, where $\ell = e$ or $\mu$.

Here we take a fresh look at the semileptonic Higgs decays  $h^0\! \to WW \! \to \ell\nu jj$
for $M_h \gtrsim 2M_W$, and  $h^0\! \to Wjj \! \to \ell\nu jj$ for $M_h \lesssim 2M_W$,
where $j$ is a hadronic jet. The overall decay rate is 6.4 times larger than 
the leptonic $WW$ mode used so far to set limits at the Tevatron, and 
at least $130$ times larger than that of the $h^0\! \to ZZ \! \to 4\ell$ ``golden mode'' 
at the LHC. Moreover,  the semileptonic $WW$ mode
is larger than any other Higgs decay mode with a triggerable lepton for $M_h \gtrsim 120$ GeV (the
decay $h^0\! \to \tau^-\tau^+$, with one of the $\tau$'s decaying leptonically, has a larger 
branching fraction for smaller $M_h$).
Thus this is a potentially interesting channel both for discovery and characterization of 
a putative Higgs resonance.

This process was first discussed as a potential Higgs discovery
channel for the SSC \cite{Stirling:1985bi,Gunion:1986cc}, emphasizing the case of
a very heavy Higgs boson, where the golden mode starts to
become rate-limited. Like the golden mode, the semileptonic  $h^0\! \to WW$ modes have the
advantage of being fully-reconstructible: when the leptonic $W$ is close to
on shell, the $W$ mass constraint determines the longitudinal momentum of
the neutrino, up to a two-fold ambiguity \cite{Gunion:1986cc}.
It is perhaps surprising then that there are currently no published Tevatron results for this channel,
and that for the LHC this channel is considered within the ATLAS and CMS physics TDRs 
only in the special case where the Higgs is produced from vector boson fusion \cite{Aad:2009wy,Ball:2007zza}.

The drawback of this channel, leading to its relative neglect in literature, is the large 
background contributions from Standard Model
processes with a leptonic $W$. These include diboson production, top quark production, and
direct inclusive $W + 2j$ production. There is also, presumably, a significant purely QCD background,
which is difficult to estimate outside of a dedicated analysis with real data. The most worrisome
background is inclusive $W + 2j$; from this background alone an ATLAS study estimates a signal
to background ratio (S/B) of $5\times 10^{-4}$, after nominal preselections \cite{Aad:2009wy}.
This is small, but not smaller than the analogous $S/B \simeq 4\times 10^{-5}$
for the $e^+e^-$ and $\mu^+\mu^-$ modes after preselection in the successful Tevatron
analyses of $h^0\! \to W^+W^- \!\to \ell^+\nu\ell^-\bar{\nu}$ \cite{D0note}.

Several approaches have been proposed for beating down these backgrounds to extract
and characterize
a signal either at the Tevatron \cite{Han:1998ma} or the LHC \cite{Aad:2009wy,Ball:2007zza,Iordanidis:1997vs}. 
For the $t\bar{t}$ background it is at least
plausible that a veto on extra hard jets, perhaps combined with a $b$-jet veto, will
give the required rejection without sacrificing much of the signal \cite{Han:1998ma,Aad:2009wy}.
For the diboson and $W + 2j$ backgrounds, the most-cited strategy is to restrict
the analysis to Higgs production via vector boson fusion \cite{Stirling:1985bi,Iordanidis:1997vs};
the requirement of forward jet tagging then gives a factor of $\sim 100$ reduction
in these backgrounds. However the reduction in the Higgs signal, versus inclusive
Higgs production, is also severe: a factor of $\sim 10$ at the Tevatron, and also at the
LHC for $M_h \lesssim 500$ GeV. A Tevatron study, looking at the similar trade-off
for the leptonic $WW$ channel, concluded that the
overall sensitivity does not improve by restricting to  vector boson fusion Higgs versus inclusive Higgs
production \cite{Mellado:2007fb}; a comparable analysis does not exist for the
semileptonic channel.

Returning to inclusive Higgs production, the remaining approaches to
background reduction and signal characterization involve angular distributions or correlations and
kinematic properties of the events, attempting to exploit the fact that the
signal involves the
production and decay of a spin zero $CP$ even resonance \cite{Han:1998ma,Stirling:1985bi,Gunion:1986cc}.

Our purpose is to revisit these approaches in a systematic way.
In this paper we examine the fully differential decay width
for the signal. Taking advantage of the fact that the 
$h^0\! \to WW \! \to \ell\nu jj$
and $h^0\! \to Wjj \to \ell\nu jj$ channels are fully reconstructible,
it is possible to make this study analytically, even
with the inclusion of some important cuts. This study is collider independent, and can
prove useful not only for the Tevatron and LHC, but also for future colliders.
A comparison with the backgrounds is very important, but is highly collider dependent;
a study of the semileptonic Higgs decays at the Tevatron 
will appear in a separate publication \cite{DobrescuLykkenWinter2010}.

In Section 2 we discuss a set of 5 kinematic variables that describe completely the
$\ell\nu jj$ final state, and then derive analytically the fully differential Higgs width.
This result allows us to identify the region of large signal in the 5-dimensional kinematic space;
we then discuss (in Section 3) some cuts that reduce efficiently the phase space for 
a given reduction in the signal. Our concluding comments are collected in Section 4.

\smallskip

\section{Fully differential width for Higgs decays to $\ell\nu jj$} \setcounter{equation}{0}

\begin{figure}[t]
\begin{center}
\unitlength=1 pt
\SetScale{0.7}\SetWidth{0.75}      
\begin{picture}(100,100)(50,-50)
\thicklines
\DashLine(-10,0)(80,0){6} 
\Photon(80,0)(150,50){3}{6}
\Line(150,50)(230,70)
\Line(150,50)(230,30)
%
\Photon(80,0)(150,-50){3}{6}
\Line(150,-50)(230,-70)
\Line(150,-50)(230,-30)
\put(14,7){$h^0$}
\put(63,25){$W^{(*)}$}
\put(60,-33){$W^{(*)}$}
\put(170,49){$\nu$}
\put(170,17){$e^+ \! , \, \mu^+$}  
\put(170,-23){$\bar{u} \; (\bar{c}) $}
\put(170,-52){$d \; (s)$}
\end{picture}
\end{center}
\vspace*{-0.2cm}
\caption{Higgs decay to a semileptonic $WW$ pair.}
\label{fig:decay}
\end{figure}
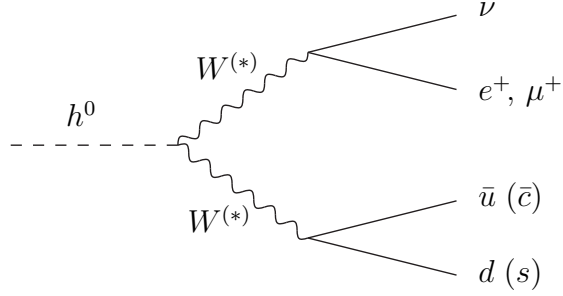

We are interested in the decay of the standard model Higgs boson into 
a pair of $W$ bosons, with one of them decaying leptonically and the other one 
hadronically. For concreteness, let us consider first the 
cascade decay $h^0\! \to W^{(*)}W^{(*)} \!\to \ell^+ \nu d \bar{u} $,
where $W^{(*)}$ represents a $W$ boson which is either on-shell or off-shell 
(in the latter case it is usually denoted by $W^*$), and $\ell$ is either an electron or a 
muon. At tree-level  in the unitary gauge there is
a single diagram, shown in Figure \ref{fig:decay}, but in practice it is easier to use the 
't Hooft-Feynman gauge because the additional diagrams due to Goldstone 
boson exchange vanish in the limit of massless fermions.
Squaring the matrix element and summing over final spins and color, and neglecting 
the fermion masses, we find a remarkably simple Lorentz-invariant result:
\be
\left| {\cal M} (h^0\! \to \ell^+\!  \nu d \bar{u} )\right|^2 = \frac{24 g^6 M_W^2  \, 
(p_u \!\cdot\! p_\ell) (p_d \!\cdot\! p_\nu)}
{ \left[\left(2 p_u \!\cdot\! p_d \!-\! M_W^2 \right)^2\! + \! M_W^2 \Gamma_W^2\right]\! 
  \left[\left(2 p_\ell \!\cdot\! p_\nu \!-\! M_W^2 \right)^2\! +\!  M_W^2 \Gamma_W^2\right] } ~~,
\ee
where we have summed over $\ell = e, \mu$.
The structure of the 4-momenta contractions in the numerator agrees with that derived in Ref.~\cite{Barger:1990mn}.

The Lorentz-invariant 4-body phase space may be written as
\be
d \Phi_4 = \frac{1}{4\pi^2} dm_{\ell\nu}^2 d m_{jj}^2 \, d \Phi_2(h^0\! \to W^{(*)+}W^{(*)-})
\,\, d\Phi_2(W^{(*)+}\!\!  \to \ell^+\! \nu) \,\, d \Phi_2(W^{(*)-}\!\! \to d\bar{u})  ~~,
\ee
where $m_{jj}$ and $m_{\ell\nu}$ are the invariant masses of the $d\bar{u}$ (the `dijet') 
and $\ell^+\!\nu$ systems, respectively:
\be
m_{jj} = \sqrt{2 p_u \!\cdot\! p_d} ~~, \;\; m_{\ell\nu} = \sqrt{2 p_\ell \!\cdot\! p_\nu} ~~.
\ee
Taking advantage of the fact that $h^0$ has spin 0, we integrate over the 
solid angles of the  $W^{(*)}$'s, and obtain the following 
two-body phase space for the $h^0 \to W^{(*)+}W^{(*)-}$ process in the Higgs 
rest frame:
\be
d \Phi_2(h^0\! \to W^{(*)+}W^{(*)-}) = \frac{1}{8\pi M_h^2 }
\left[ \left(M_h^2 + m_{\ell\nu}^2 - m_{jj}^2 \right)^2 - 4 m_{\ell\nu}^2 M_h^2  \right]^{\! 1/2}  ~~.
\ee

For the decays of each (virtual) $W^{(*)}$ boson, it is convenient to write 
the two-body phase space (which is also Lorentz invariant) in the respective rest frame.
Let us discuss what kinematic variables are convenient to use in the two $W^{(*)}$ rest frames.
The 4-momenta $p_u$, $p_d$, $p_\ell$ and $p_\nu$ are constrained by energy-momentum conservation
such that in the rest frame of the Higgs boson
the fully differential width for this decay depends on five kinematic variables. 
Two of these may be taken to be $m_{jj}$ and $m_{\ell\nu}$
within the following ranges allowed by energy-momentum conservation:
\be
0 \le m_{jj} \le M_h - m_{\ell\nu}  ~~, \;\; 0 \le m_{\ell\nu} \le M_h  ~~.
\ee
Two more kinematic variables may be taken to be the polar ($\theta_\ell^0$)  and 
azimuthal ($\varphi_\ell$) angles between the charged lepton momentum 
$\vec{p}_\ell^{\; 0}$ in the $\ell\nu$ rest frame
and the $W^{(*)+}$ momentum in the Higgs rest frame, which is given by $-(\vec{p}_u + \vec{p}_d)$.
The two-body phase space for the $W^{(*)+}\!\!  \to \ell^+\!\nu$ decay is given by
\be
d \Phi_2(W^{(*)+}\!\!  \to \ell^+\!\nu) = \frac{1}{32\pi^2} \; 
d \varphi_\ell \; d (\cos\theta_\ell^0 ) ~.
\ee

The remaining kinematic variable must describe the motion of the $d\bar{u}$ system. 
When choosing it, one should take into account that the $d$ quark and $\bar{u}$ antiquark 
hadronize giving rise to jets. 
Since it is practically impossible to tell whether a jet originates from the 
quark or the antiquark, we have to define a kinematic variable which is
independent of the jet origin. 
In the dijet rest frame, it is convenient to use the angle $\theta_{j}^0$  between 
$-(\vec{p}_\ell + \vec{p}_\nu)$ 
and the momentum of the jet which in the Higgs rest frame has the highest energy (we refer to this as the 
`leading' jet, and we label it by $j_1$). 

\begin{figure}[t]
\begin{center}
\unitlength=1 pt
\SetScale{1}\SetWidth{0.5}      
\begin{picture}(300,130)(-150,-50)
\thinlines
\put(-140,0){\vector(1,0){310}} \put(164,-12){$z$}
\put(0,-58){\vector(0,1){135}} \put(-13,65){$x$}
\put(30,30){\vector(-1,-1){91}} \put(-45,-57){$y$}
\thicklines
\LongArrow(0,0)(71,42) \put(80,38){$\vec{p}_\ell$}
\Line(0.5,-0.5)(71.5,41.5)
\Line(-0.5,0.5)(70.5,42.5)
\ArrowArc(0,-8)(55,8,39)\put(57,12){$\theta_\ell$}
\LongArrow(0,0)(-97,27) \put(-104,38){$\vec{p}_{j_1}$}
\Line(0.5,0.5)(-96.5,27.5)
\Line(-0.5,-0.5)(-97.5,26.5)
\ArrowArcn(0,0)(75,180,164)\put(-88,7){$\theta_j$}
\LongArrow(0,0)(-73,-27) \put(-67,-37){$\vec{p}_{j_2}$}
\Line(0.5,-0.5)(-72.5,-27.5)
\Line(-0.5,0.5)(-73.5,-26.5)
\DashLine(0,0)(56,-41){3} 
\LongArrow(52,-39)(56,-41) \put(54,-31){$\vec{p}_\nu$}
\DashLine(0.5,0.5)(56.5,-40.5){3}
\DashLine(-0.5,-0.5)(55.5,-41.5){3}
\DashLine(-36,43)(70,43){4} 
\DashLine(-36,43)(0,0){3} 
\LongArrow(-33,39)(-36,43) \put(-59,31){$(\vec{p}_\ell)_{xy}$}
\ArrowArc(0,12)(11,90,170 )\put(-16,27){$\varphi_\ell$}
\end{picture}
\end{center}
\caption{Angular variables describing the $h^0\! \to W^{(*)}W^{(*)} \!\to \ell \nu jj $ decay,
in the Higgs rest frame.
The two jets are in the $zx$ plane. The $x$ component of the leading jet ($j_1$) is always positive.  
The angle between the $x$ axis and the 
projection of $\vec{p}_\ell$ onto the $xy$ plane is $0 \le \varphi_\ell < 2\pi$.
Boosting along the $z$ axis to the rest frame of the $\vec{p}_\ell + \vec{p}_\nu$ system, the angle $\theta_\ell$
becomes $\theta_\ell^0$. Boosting to the dijet rest frame, the angle $\theta_j$
becomes $\theta_j^0$.
}
\label{fig:axes}
\end{figure}
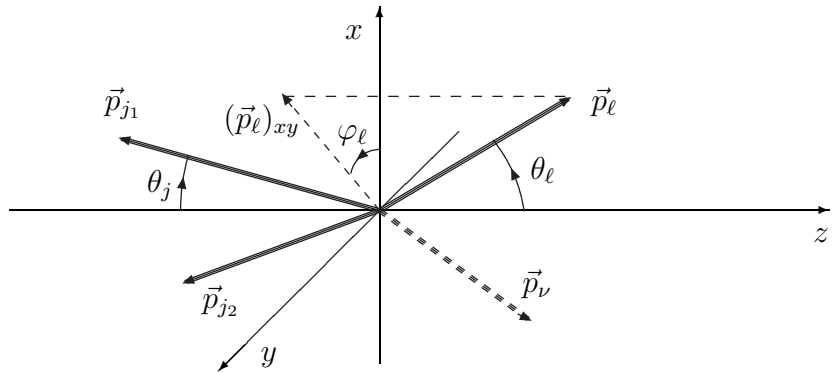

To be precise, let us define the $z$ axis along the  3-momentum of the $\ell\nu$ system
in the $h^0$ rest frame, and the $x$ axis such that the dijet plane is 
$zx$ and the $x$-component of the $j_1$ jet momentum is positive; 
the $y$ axis is perpendicular to the dijet plane.
We then boost along the $z$ axis to the $\ell\nu$ rest frame where the charged lepton and neutrino 
are back to back. Then $\theta_\ell^0$ is the angle between 
$\vec{p}_\ell^{\; 0}$ and the $z$ axis, while $\varphi_\ell$ is the angle between the 
$x$ axis and the projection of $\vec{p}_\ell^{\; 0}$ onto the $xy$ plane 
(see Figure \ref{fig:axes}). 
Note that $\varphi_\ell$ is invariant under boosts along the $z$ axis, so it can equivalently 
be defined (up to a two-fold ambiguity) as the angle between the dijet plane and the $\ell\nu$ plane
in the Higgs rest frame. 
Finally, $\theta_{j}^0$ is the angle in the dijet rest frame between the $-z$ axis and the 
jet momentum whose $z$ component is negative (this is $j_1$). 
The physical ranges for these angles are
\be
0 \le \theta_\ell^0 \le \pi \;\;\;\; ,  \;\;\;\; 
0 \le \varphi_\ell < 2 \pi \;\;\;\; ,  \;\;\;\; 
0 \le \theta_{j}^0 \le \frac{\pi}{2}   \;\; .
\label{ranges}
\ee

The two-body phase space for the $W^{(*)-}\!\! \to jj$ decay, after integrating over the 
azimuthal angle of the dijet plane (which is not observable), is 
\be
d \Phi_2(W^{(*)+}\!\!  \to d \bar{u} ) = \frac{1}{16\pi} \; 
d (\cos\theta_j^0 ) \sum_{j_1 = \bar{u}, d}  \delta_{j_1q} ~~,
\ee
where the sum over the  identity of the leading jet resolves the ambiguity caused 
by the inability to measure which jet originates from $\bar{u}$. 

In terms of the five kinematic variables defined above 
($m_{\ell\nu},m_{jj},\theta_\ell^0,\theta_{j}^0, \varphi_\ell$), we find
that if the leading jet in the Higgs rest frame originates from the $\bar{u}$ parton, then
\be
(p_u \!\cdot\! p_\ell) (p_d \!\cdot\! p_\nu) =
\frac{1}{16} m_{\ell\nu}^2 m_{jj}^2
\left[ \left( \gamma_a(1 +  c_j c_\ell )  -  s_j s_\ell \cos\varphi_\ell \right)^2 
-  \left( \gamma_a^2 - 1 \right) \left( c_j + c_\ell \right)^{\! 2}  \rule{0mm}{3.9mm} \right]
~~,
\label{pupd}
\ee
while if the leading jet originates from the $d$ parton then in the above 
equation one should make the following substitutions: $c_j \to -c_j$ and  $s_j \to -s_j$. 
We introduced here a simplifying notation:
\be
c_{\ell,j} \equiv \cos\theta_{\ell,j}^0 \;\;  ~, \;\;  s_{\ell,j} \equiv \sin\theta_{\ell,j}^0 ~,
\ee
and also
\be
\gamma_a \equiv \gamma_a(m_{\ell\nu},m_{jj}) = \frac{M_h^2 - m_{\ell\nu}^2 - m_{jj}^2}{2m_{\ell\nu} m_{jj}} \, \ge 1 ~~.
\ee
The latter quantity is related to the rapidity distributions of the $\ell\nu$ and dijet systems by
$\gamma_a = \cosh (y_{\ell\nu} - y_{jj})$.
Note that $\gamma_a =1$ corresponds to the case where the dijet and $\ell\nu$ systems are at rest in the Higgs rest frame. 
The angular distribution given in Eq.~(\ref{pupd}) is consistent with the results given in 
Ref.~\cite{Barger:1993wt,Buszello:2002uu}.

The fully differential width for the Higgs decay into $\ell\nu jj$ is 
\be
d \Gamma (h^0\! \to \ell \nu jj) = \frac{2}{M_h} d \Phi_4 
\left| {\cal M} (h^0 \to \ell^+\! \nu d \bar{u})\right|^2   ~~,
\ee
where we have included a factor of 4 to take into account the $s\bar{c}$ quark-antiquark pair
(the decays involving $b$ jets are negligible) and also the decay where the lepton charge is negative.
Putting everything together, and summing over the identity of the leading jet, we obtain the quintuply
differential Higgs width:
\be
\hspace*{-0.9em} 
\frac{d\Gamma (h^0\! \to \ell \nu jj) }{d m_{\ell\nu}^2 d m_{jj}^2  d c_\ell d c_j d \varphi_\ell }
= \frac{3 g^6 M_W^2 }{(4\pi)^6 M_h^3} 
\, \frac{m_{\ell\nu}^3m_{jj}^3 \left(\gamma_a^2-1 \right)^{\! 1/2} 
f(\gamma_a,\theta_\ell^0,\theta_j^0,\varphi_\ell) }
{\left[\left( m_{jj}^2 \!-\! M_W^2 \right)^2\! + \! M_W^2 \Gamma_W^2\right] \!
  \left[\left( m_{\ell\nu}^2 \!-\! M_W^2 \right)^2\! + \! M_W^2 \Gamma_W^2\right] } 
\label{quintuple}
\ee
where the angular dependence enters entirely through
\be
f(\gamma_a,\theta_\ell^0,\theta_{j}^0,\varphi_\ell) =
\left( \gamma_a c_\ell c_j - s_\ell s_j \cos\varphi_\ell \right)^2 \!
- (\gamma_a^2 - 1) \left( c_j^2  + c_\ell^2 \right) + \gamma_a^2  ~~.
\label{function}
\ee

The amplitude for $h^0$ decay into longitudinally-polarized $W$'s is proportional to 
$\gamma_a$.
Collecting just the terms  proportional to $\gamma_a^2$ in (\ref{function}), one obtains
$\gamma_a^2 \, s_\ell^2 s_j^2$, which is 
the contribution from purely longitudinal $W$'s.
In the heavy Higgs limit this behavior dominates, but for $M_h \lesssim 200$ GeV
we have $\gamma_a \lesssim 2$, and we get important contributions
from both the longitudinal and transverse $W$ polarizations, as well as their interference 
which is the cross term proportional to $\gamma_a$ in (\ref{function}).

\smallskip
\section{Signal-friendly cuts} 

Having derived the quintuply-differential width in Eq.~(\ref{quintuple}), one can now 
design a set of cuts that preserve as much as possible of the signal for a given reduction
in the space spanned by the five kinematic variables.

\subsection{Azimuthal angle}

Let us first discuss the $\varphi_\ell$ dependence. 
The  differential width is maximized when $\cos\varphi_\ell = 1$ and $c_\ell \le 0$,
or  $\cos\varphi_\ell = -1$ and $c_\ell \ge 0$ (note that $s_\ell,s_j,c_j \ge 0$).
This means that the $\ell\nu$ plane and the dijet plane tend to be aligned.
We therefore impose a cut
\bear
0 \le \varphi_\ell < \eta_\varphi \pi \;\;\;  {\rm or} \;\; \; 
\left(2 - \eta_\varphi \right)\pi \le \varphi_\ell \le 2\pi  ~~~, \;\; 
{\rm if} \;\;  c_\ell \le 0 ~~;
\nonumber \\ [3mm] 
\left(1-\eta_\varphi \right)\pi < \varphi_\ell < \left(1+\eta_\varphi \right)\pi 
\;\;\;\;  ~~~, \;\; 
{\rm if} \;\;  c_\ell \ge 0 ~~.
\label{azimuth}
\eear
Here $0< \eta_\varphi \leq 1$ is a cut parameter. 
For $ \eta_\varphi = 1$ there is no cut, while for  $ \eta_\varphi \to 0$ the whole phase space is cut.
Integrating over the above $\varphi_\ell$ range we get 
\be
f_{\eta_\varphi}(\gamma_a, c_\ell, c_j) \equiv \int \! d\varphi_\ell \,  
f(\gamma_a,\theta_\ell^0,\theta_{j}^0,\varphi_\ell) 
= f_0 + 2 \eta_\varphi\pi \gamma_a^2  s_\ell^2 s_j^2 + 
4    \sin\!\left(\eta_\varphi \pi \right) \gamma_a s_\ell |c_\ell| \, s_j c_j  ~,
\label{eq:feta}
\ee
where
\be
f_0 = 2 \eta_\varphi\pi \! \left(c_j^2 + c_\ell^2 \right)
+ \left(\eta_\varphi\pi + \frac{\sin (2 \eta_\varphi \pi)}{2}    \right)
\! s_\ell^2 s_j^2 
\ee

The range of variables shown in Eq.~(\ref{ranges}) gives $-1 \le c_\ell \le 1$ 
and $0 \le c_j \le 1$. However, $f_{\eta_\varphi}(\gamma_a, c_\ell, c_j)$ is symmetric under 
$c_\ell \to -c_\ell$, so that it is sufficient to take $0 \le c_\ell \le 1$ and to include 
a factor of 2 in the integral over $c_\ell$.

In general, $\eta_\varphi$ may be chosen to be some function of the other four 
kinematic variables, with the aim of optimizing the cuts. In what follows we 
will ignore this refinement.

\subsection{Invariant masses}

It is convenient to split the range of $M_h$ in three regions: above, near or below
the $WW$ threshold. 

\subsubsection{Above the $WW$ threshold}

In this subsection we analyze the `above threshold' region, which we define as
\be
M_h - 2 M_W \gg \frac{\Gamma_W}{2\pi}  ~.
\ee
Given that $\Gamma_W \approx 2.1$ GeV, the above threshold region
is $M_h \gtrsim 165$ GeV.

In this region, the narrow width approximation works very well.
The vast majority of the signal is concentrated around $m_{jj} \simeq m_{\ell\nu}  \simeq M_W$, and thus it is 
convenient to impose the cuts 
\be
1 - \delta_\ell \le \frac{m_{\ell\nu}}{M_W} \le 1 + \delta_\ell 
\;\;\; , \;\;\;
1 - \delta_j \le \frac{m_{jj}}{M_W} \le 1 + \delta_j   ~~,
\label{abovecuts}
\ee
In order to preserve most of the signal while cutting efficiently the phase space,
the window parameters $\delta_j$ and $\delta_\ell$ must satisfy 
\be
\frac{\Gamma_W}{\pi M_W} \ll \delta_{\ell , j} \ll 1  ~~.
\label{deltas}
\ee
Values for $\delta_{\ell , j}$ in the $5-10$\% range satisfy the above requirements. 
In these circumstances, we may generically express the narrow width approximation as
\be
\int_{M_W(1 - \delta)}^{M_W(1 + \delta) }\!\! dm \,
\frac{m \, F(m)}{\left( m^2 \!-\! M_W^2 \right)^2\! + \! M_W^2 \Gamma_W^2} 
\approx \frac{\pi}{2 M_W \Gamma_W} F(M_W) \left[ 1 + O\left(\frac{\Gamma_W}{\pi M_W \delta_{\ell, j} }\right)\right]  ~~,
\ee
where $F(m)$ is an arbitrary nonsingular function.
Note that the size of the windows ($2\delta_{\ell, j}$) affects 
only the small corrections as long as Eq.~(\ref{deltas}) is satisfied.

Applying this formula to the differential Higgs width, we find
\be
\hspace*{-0.9em} 
\frac{d \Gamma (h^0\! \to \ell \nu jj) }{ d c_\ell d c_j  }
= \frac{3 g^6 }{(8\pi)^4}\frac{M_W^4}{M_h \Gamma_W^2} 
\, \left(1 - \frac{4M_W^2}{M_h^2}\right)^{\! 1/2} 
f_{\eta_\varphi}(\bar{\gamma}_a,c_\ell, c_j)  ~,
\label{double}
\ee
with $0\le c_\ell,c_j \le 1$ and 
\be
\bar{\gamma}_a = \frac{M_h^2}{2M_W^2} - 1 ~~.
\ee

\subsubsection{Near the $WW$ threshold}

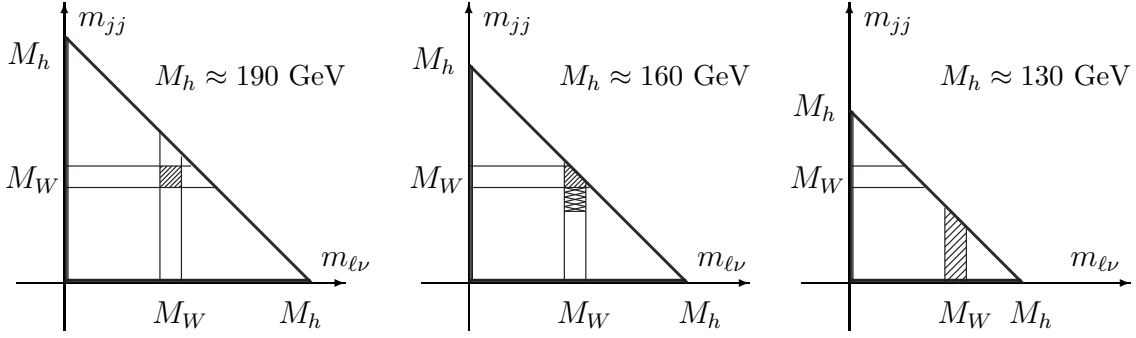
\begin{figure}[t]
\begin{center}
\unitlength=0.9 pt
\SetScale{0.9}\SetWidth{0.5}      
\begin{picture}(300,140)(-130,0)
\put(-200,0){  \put(38,83){\small $M_h \approx 190$ GeV}
\thinlines
\put(-20,0){\vector(1,0){138}} \put(108,7){$m_{\ell\nu}$} \put(90,-17){$M_h$} \put(37,-17){$M_W$}
\put(0,-20){\vector(0,1){138}} \put(6,108){$m_{jj}$}\put(-24,92){$M_h$}\put(-25,40){$M_W$}
\thicklines
\Line(40,0)(40,64) \Line(49,0)(49,53)  
\Line(0,40)(64,40) \Line(0,49)(53,49) 
\Line(0,103.5)(103.5,0)\Line(0,103)(103,0)\Line(0,104)(104,0)  
\Line(0,0.5)(103,0.5)\Line(0,1)(103,1)\Line(0,1.5)(103,1.5)
\Line(0.5,0)(0.5,103)\Line(1,0)(1,103)\Line(1.5,0)(1.5,103)
\thinlines
\Line(40,40)(49,49)
\Line(40,42)(47,49)\Line(40,44)(45,49)\Line(40,46)(43,49)\Line(40,48)(41,49)
\Line(42,40)(49,47)\Line(44,40)(49,45)\Line(46,40)(49,43)\Line(48,40)(49,41)
}
\put(-30,0){\put(38,83){\small $M_h \approx 160$ GeV}
\thinlines
\put(-20,0){\vector(1,0){138}} \put(96,7){$m_{\ell\nu}$} \put(88,-17){$M_h$} \put(37,-17){$M_W$}
\put(0,-20){\vector(0,1){138}} \put(6,108){$m_{jj}$}\put(-24,90){$M_h$}\put(-25,40){$M_W$}
\thicklines
\Line(40,0)(40,50.5) \Line(49,0)(49,42)
\Line(0,40)(50.5,40) \Line(0,49)(42,49)
\Line(0,91.5)(91.5,0)\Line(0,91)(91,0)\Line(0,92)(92,0)  
\Line(0,0.5)(91.5,0.5)\Line(0,1)(91,1)\Line(0,1.5)(91,1.5)
\Line(0.5,0)(0.5,91.5)\Line(1,0)(1,91)\Line(1.5,0)(1.5,91)
\thinlines
\Line(40,40)(46,46)
\Line(40,42)(45,47)\Line(40,44)(44,48)\Line(40,46)(43,49)\Line(40,48)(41,49)
\Line(42,40)(47,45)\Line(44,40)(48,44)\Line(46,40)(49,43)\Line(48,40)(49,41)
\Line(40,38)(49,40)\Line(40,36)(49,38)\Line(40,34)(49,36)\Line(40,32)(49,34)\Line(40,30)(49,32)
\Line(40,40)(49,38)\Line(40,38)(49,36)\Line(40,36)(49,34)\Line(40,34)(49,32)\Line(40,32)(49,30)
\Line(40,30)(49,30)
}
\put(130,0){ \put(38,83){\small $M_h \approx 130$ GeV}
\thinlines
\put(-20,0){\vector(1,0){138}} \put(93,7){$m_{\ell\nu}$} \put(67,-17){$M_h$} \put(37,-17){$M_W$}
\put(0,-20){\vector(0,1){138}} \put(6,108){$m_{jj}$}\put(-24,70){$M_h$}\put(-25,40){$M_W$}
\thicklines
\Line(40,0)(40,31.5) \Line(49,0)(49,23)
\Line(0,40)(31.5,40) \Line(0,49)(23,49)
\Line(0,72.5)(72.5,0)\Line(0,72)(72,0)\Line(0,73)(73,0)   
\Line(0,0.5)(72,0.5)\Line(0,1)(72,1)\Line(0,1.5)(72,1.5)
\Line(0.5,0)(0.5,72)\Line(1,0)(1,72)\Line(1.5,0)(1.5,72)
\thinlines
\Line(43,0)(49,6)\Line(46,0)(49,3)
\Line(40,0)(49,9)\Line(40,3)(49,12)\Line(40,6)(49,15)
\Line(40,9)(49,18)\Line(40,12)(49,21)\Line(40,15)(49,24)
\Line(40,18)(47,25)\Line(40,21)(46,27)\Line(40,24)(44,28)\Line(40,27)(43,30)
}
\end{picture}
\end{center}
\caption{Kinematically allowed region for the dijet and $\ell\nu$ invariant masses. The three plots correspond to 
the cases of above, near, and below the WW threshold, respectively. The shaded regions are the ones kept by 
cuts given in Eqs.~(\ref{abovecuts}), (\ref{integration-limits}), and (\ref{abovecuts-near})}
\label{fig:region}
\end{figure}

Let us now analyze the `near threshold' region, which we define as
\be
\left| M_h - 2 M_W \right| \lesssim 2 \Gamma_W  ~.
\ee
This corresponds to a range for the Higgs mass of roughly  $156$ GeV $ \lesssim M_h \lesssim 165$ GeV.
The cuts shown in Eq.~(\ref{abovecuts}) need to be modified here because of the kinematical constraint
$m_{\ell\nu} + m_{jj}  \le M_h$.  The width has a sharp peak at $m_{\ell\nu} = M_W$ and 
$m_{jj}={\rm min} (M_h - M_W ; M_W)$, so that we impose
\bear
&& \left[ M_h - M_W(1 - \delta_\ell) \right]   
(1 - \eta_j) 
\le m_{jj} \le M_h - m_{\ell\nu}  ~~, 
\label{integration-limits}
\\ [4mm]
&& 1 - \delta_\ell \le  \frac{m_{\ell\nu}}{M_W} \le 1 + \delta_\ell ~~,
\label{abovecuts-near}
\eear
where $0 < \eta_j \le 1$ is a cut parameter describing the length of the vertical 
shaded region in the middle plot of Figure~\ref{fig:region}. 

With these cuts, the narrow width approximation may be used only for the $W$ 
decaying into $\ell\nu$.
The Higgs differential width may be written as 
\bear
\hspace*{-0.9em} 
\frac{d\Gamma (h^0\! \to \ell \nu jj) }{d c_\ell d c_j }
&\!\! =\!\! & \frac{3 g^6 M_W^3}{2(4\pi)^5 M_h \Gamma_W} 
\left( f_0 I_0 + 2\eta_\varphi\pi s_\ell^2 s_j^2   I_2 +   4  \sin (\eta_\varphi \pi ) s_\ell c_\ell \, s_j c_j   I_1 
\right)~,
\label{double-below}
\eear
with $0\le c_\ell,c_j \le 1$. 
Here $I_n$ with $n = 0,1,2$ are integrals over the $jj$ invariant mass:
\be
\hspace*{-0.7em}
I_n = \frac{1}{(2M_W)^n}\int 
d m_{jj} \, \frac{ m_{jj}^{3-n} \left(M_h^2 - \! M_W^2\! - m_{jj}^2\right)^n }
{\left( m_{jj}^2 \!-\! M_W^2 \right)^2\! + \! M_W^2 \Gamma_W^2 }
\left[\! \left(\! 1 \! - \! \frac{M_W^2\! - m_{jj}^2}{M_h^2}\! \right)^{\!\!2} \!\! - \frac{4 m_{jj}^2}{M_h^2} \right]^{\! 1/2} ~,
\label{integrals}
\ee
where the integration limits are shown in Eq.~(\ref{integration-limits}).
The cut parameter may be chosen to grow from $\eta_j = 0$ for $M_h \approx 165$ GeV 
(most of the signal is concentrated in the intersection of the two bands shown in the middle plot
of Figure~\ref{fig:region}),
to $\eta_j = O(10\%)$ for $M_h \approx 156$ GeV (because the tail of the invariant mass 
distribution grows larger for smaller $M_h$). 

Besides the range of invariant masses defined by Eq.~(\ref{abovecuts-near}), there is a range 
around $m_{jj} \simeq M_W$ with $m_{\ell\nu}$ below but close to $
(1 - \delta_\ell)M_W$ where the signal is relatively large. 
There, however, the reconstruction of the Higgs peak 
is no longer possible because of the unknown neutrino momentum along the beam axis, and therefore 
we do not include its contribution in the $I_n$ integrals.

Note that Eq.~(\ref{double-below}) is valid both near and above the threshold; using 
the narrow width approximation for $m_{jj}$ in Eq.~(\ref{integrals}) one recovers 
the result (\ref{double}) obtained for the region  above the threshold. 

\subsubsection{Below the $WW$ threshold}

Finally, there is the `below threshold' region, which we define as
\be
2 M_W - M_h \gg \frac{\Gamma_W}{2\pi}  ~,
\ee
and for practical purposes may be taken to be $M_h \lesssim 156$ GeV.
In this region also only one of the $W$ propagators may be treated in the narrow 
width approximation (at least for $M_h \gtrsim 130$ GeV). 

We consider here only the case where $m_{\ell\nu}  \simeq M_W$ because the other region
of large signal ($m_{jj} \simeq M_W$) does not allow the Higgs reconstruction.
The dijet invariant mass then satisfies $m_{jj} \le M_h - M_W$.
The differential Higgs width is then given by 
Eq.~(\ref{double-below}), with the integration limits in Eq.~(\ref{integration-limits})
now being well approximated by 
\be
\left( M_h - M_W \right) (1 - \eta_j) \le m_{jj} \le M_h - M_W ~~.
\ee
The cut parameter here needs to be larger than in the `near threshold' region because 
the signal is no longer so sharply peaked. In fact, for $M_h \lesssim 140$ GeV, 
the slope of the $m_{jj}$ distribution is so flat that it is preferrable not to cut 
it ({\it i.e.}, $ \eta_j = 1$).

\subsection{Polar angles}

\begin{figure}[t]\center
\psfrag{bb}[t]{$\; b$}
\psfrag{Mh}[B]{\small $M_h$ [GeV]}
\psfig{file=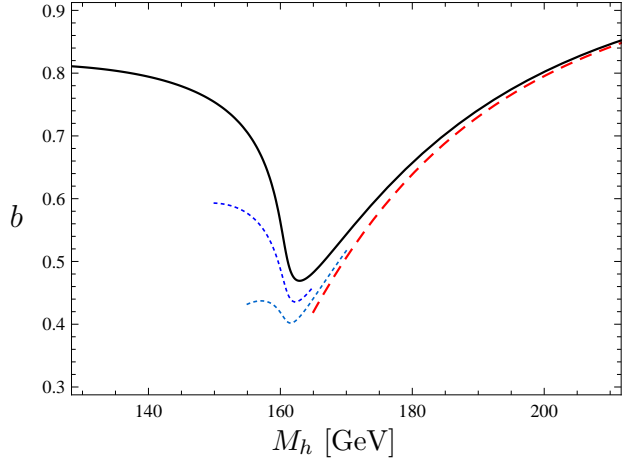,width=8.1cm,angle=0}\hspace*{0.9cm}
\caption{$M_h$ dependence of the parameter $b$ that controls the behavior 
of $\hat{f}_1(c_\ell, c_j)$. The solid line involves the narrow width approximation only for 
$W \to \ell\nu$, as in Eq.~(\ref{bbelow}), and no $m_{jj}$ cut.
The bottom (top) dotted line represents the values of $b$ for a cut $m_{jj} > 65$ (40) GeV.
The dashed line involves the narrow width 
approximation for both $W$ bosons, as in Eq.~(\ref{babove}), with no  $m_{jj}$ cut.
}
\label{fig:bplot}
\end{figure}

Let us consider only the simple cases where $\eta_\varphi = 1$ (no cut on $\varphi$)
and $\eta_\varphi = 1/2$.
For $\eta_\varphi = 1$ the interference term vanishes, and the dependence of the Higgs decay width on $c_\ell$
and $c_j$ is entirely contained in the following function:
\be
\hat{f}_1(c_\ell, c_j) = 1 + c_\ell^2c_j^2 - b \left(c_\ell^2 + c_j^2\right) ~~,
\label{fhat1}
\ee
where $b$ is a parameter that depends on  $M_h$ and on the invariant mass cut 
via the integrals $I_0$ and $I_2$ given in Eq.~(\ref{integrals}):
\be
b = \frac{2 I_2 - I_0}{2I_2 + I_0}  ~~.
\label{bbelow}
\ee
Above the threshold, where we can use the narrow width approximation for the 
$W$ boson decaying to jets, the parameter $b$ depends only on $M_h$:
\be
b \simeq 1 - \frac{4M_W^4}{M_h^4 - 4 M_h^2 M_W^2 + 6M_W^4 }  \;\; , \;\; {\rm for} \, \; 
M_h \gtrsim 165 \; {\rm GeV} ~,
\label{babove}
\ee
This parameter varies from $b\approx 0.47$ for $M_h = 163$ GeV to $b\approx 0.8$ for $M_h = 135$ GeV
or $M_h = 200$ GeV (see Figure~\ref{fig:bplot}), and grows monotonically towards the asymptotic value
$b=1$ in the $M_h \gg 2M_W$ limit. Near or below threshold, the values of $b$ decrease 
in the presence of a $m_{jj}$ cut.

\begin{figure}[t]\center
\psfrag{cj}[b]{$c_\ell$}\psfrag{cl}[t]{$c_j$}
\hspace*{-0.9cm}
\psfig{file=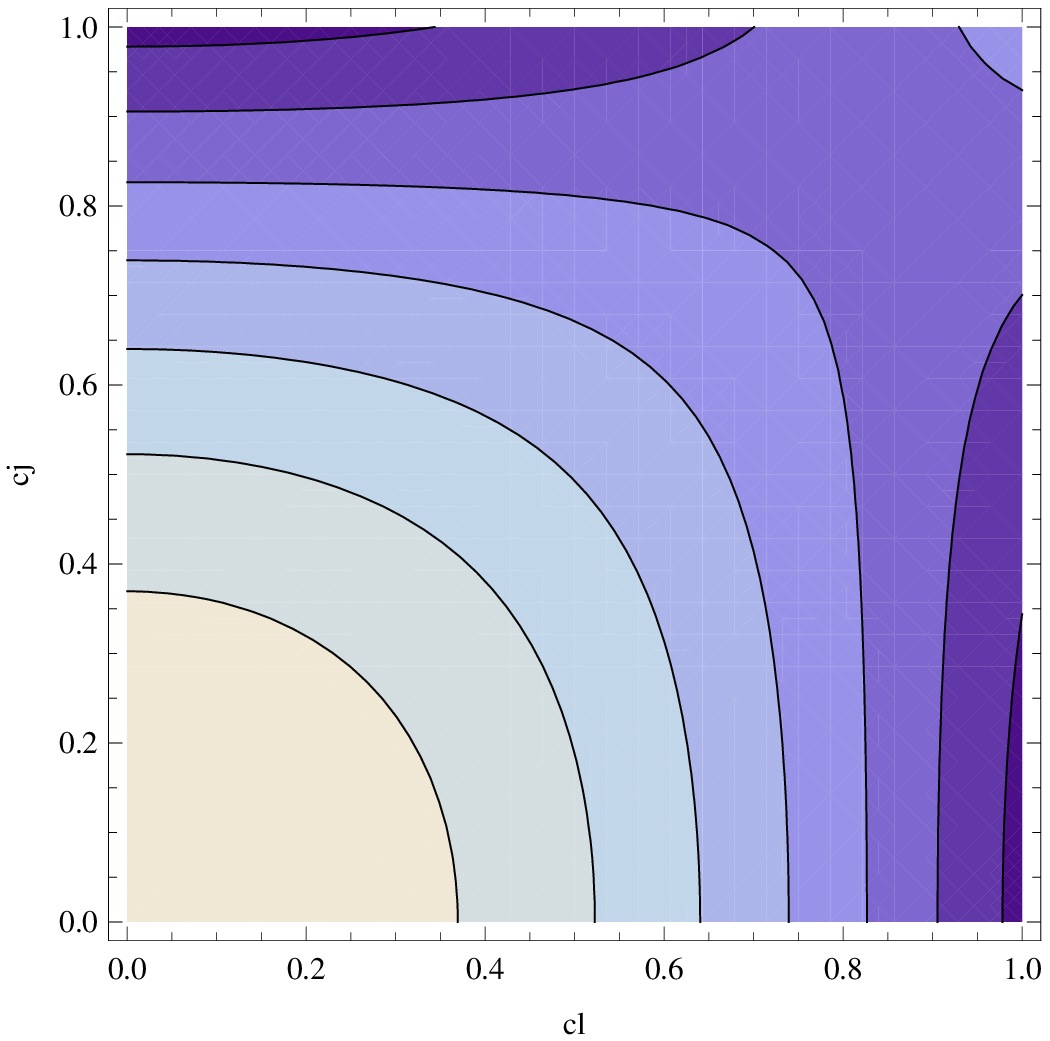,width=7cm,angle=0}\hspace*{0.9cm}
\psfig{file=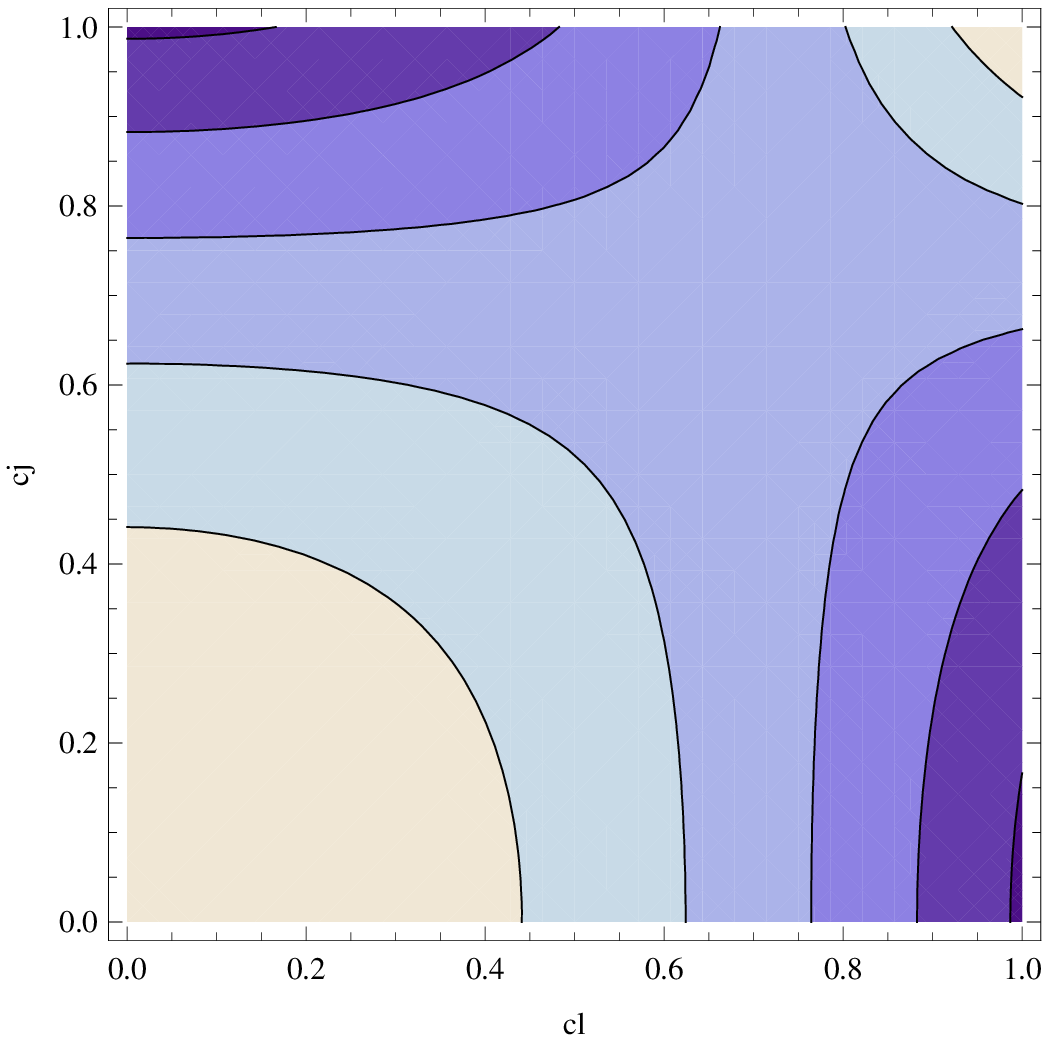,width=7cm,angle=0}
\caption{Contour lines of the $\hat{f}_1(c_\ell, c_j)$ function defined in Eq.~(\ref{fhat1}). 
Left panel: $b=0.73$, corresponding to $M_h = 190$ GeV and $m_{jj} > 70$ GeV, or $M_h \approx 151$ GeV and no $m_{jj}$ cut.
Right panel: $b=0.51$, corresponding to 
$M_h = 170$ GeV and $m_{jj} > 70$ GeV, or $M_h \approx 160$ GeV and  $m_{jj} > 40$ GeV. 
From the maximum at (0,0) where $\hat{f}_1=1$, each contour line marks a change in $\hat{f}_1$ of 0.1;
the lighter shades represent higher values. It is useful to cut along a
contour line close to $c_\ell = c_j = 0$. For 160 GeV $\lesssim M_h \lesssim 170$ GeV, 
one should also cut along a contour line close to $c_\ell = c_j = 1$. }
\label{fig:ccplot}
\end{figure}

The function $\hat{f}_1(c_\ell, c_j)$, with $0\le c_\ell,c_j \le 1$,
has only two local maxima: a broad peak at  $c_\ell=c_j=0$ (where $\hat{f}_1 = 1$), 
and a narrow peak at $c_\ell=c_j=1$ (where $\hat{f}_1 = 2-2b$).
The highest maximum  is the one at (0,0) for $b \ge 1/2$,
and at $(1,1)$ for $b \le 1/2$ (this corresponds to 
$161 \le M_h\le 166$ GeV if there is no $m_{jj}$ cut).
In between these peaks there is a saddle point at $c_\ell=c_j=\sqrt{b}$. 
There are two minima (where $\hat{f}_1 = 1-b$) at $c_\ell=0, c_j=1$ and 
$c_\ell=1, c_j=0$, as shown in Figure~\ref{fig:ccplot}.

A cut that preserves as much as possible of the signal 
for a given reduction of the allowed $(c_\ell,c_j)$ region
must include the broad peak, so that it can be parametrized as
\be
\hat{f}_1(c_\ell, c_j) \ge 1 - \eta_c b^2 ~,
\label{contour-cut}
\ee
where $0 < \eta_c \le 1$ is a parameter describing how large $f_1$ is along the cut. 
For  $\eta_c = 1$ the broad peak is included up to the height of the saddle point; for $\eta_c = 0$ the whole broad peak is cut.
The region of integration corresponding to this cut is
\be
0 \le c_\ell^2 \le b  \, \frac{\eta_c b - c_j^2}{b - c_j^2}
\;\; {\rm for} \; \; 0\le c_j^2  < \eta_c b ~~.
\ee
For  $160 \lesssim M_h \lesssim 170$, the sharper peak at  $c_\ell = c_j = 1$ is high enough to warrant an additional
cut around it: 
\be
1 \ge c_\ell^2 \ge b \, \frac{c_j^2 - b \eta_c}{c_j^2 - b}
\;\; {\rm for} \; \; 1 \ge c_j^2 > b  \frac{1 - \eta_c b}{1-b}  ~,
\ee
where Eq.~(\ref{contour-cut}) has been used again.
Integrating over $c_\ell$ and $c_j$ we find
\bear
\hspace*{-0.9em} 
\Gamma (h^0\! \to \ell \nu jj) 
&\!\! =\!\! & \frac{3 g^6 M_W^3}{8 (4\pi)^4 M_h \Gamma_W} \left( I_0 + 2 I_2 \right)
\int d c_j d c_\ell \hat{f}_1 ~~.
\label{width-cuts}
\eear

\begin{figure}[t]\center
\psfrag{cj}[b]{$c_\ell$}\psfrag{cl}[t]{$c_j$}
\hspace*{-0.9cm}
\psfig{file=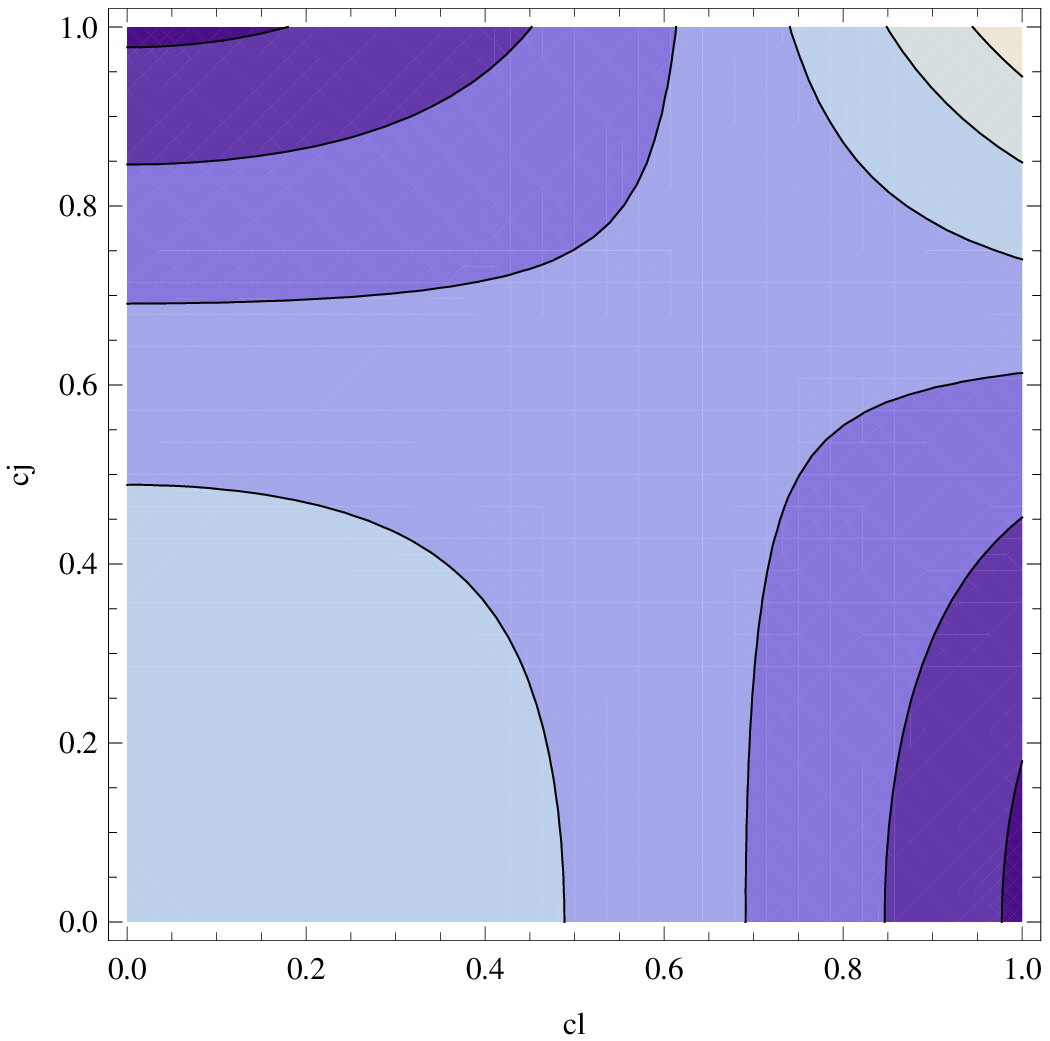,width=7cm,angle=0}\hspace*{0.9cm}
\psfig{file=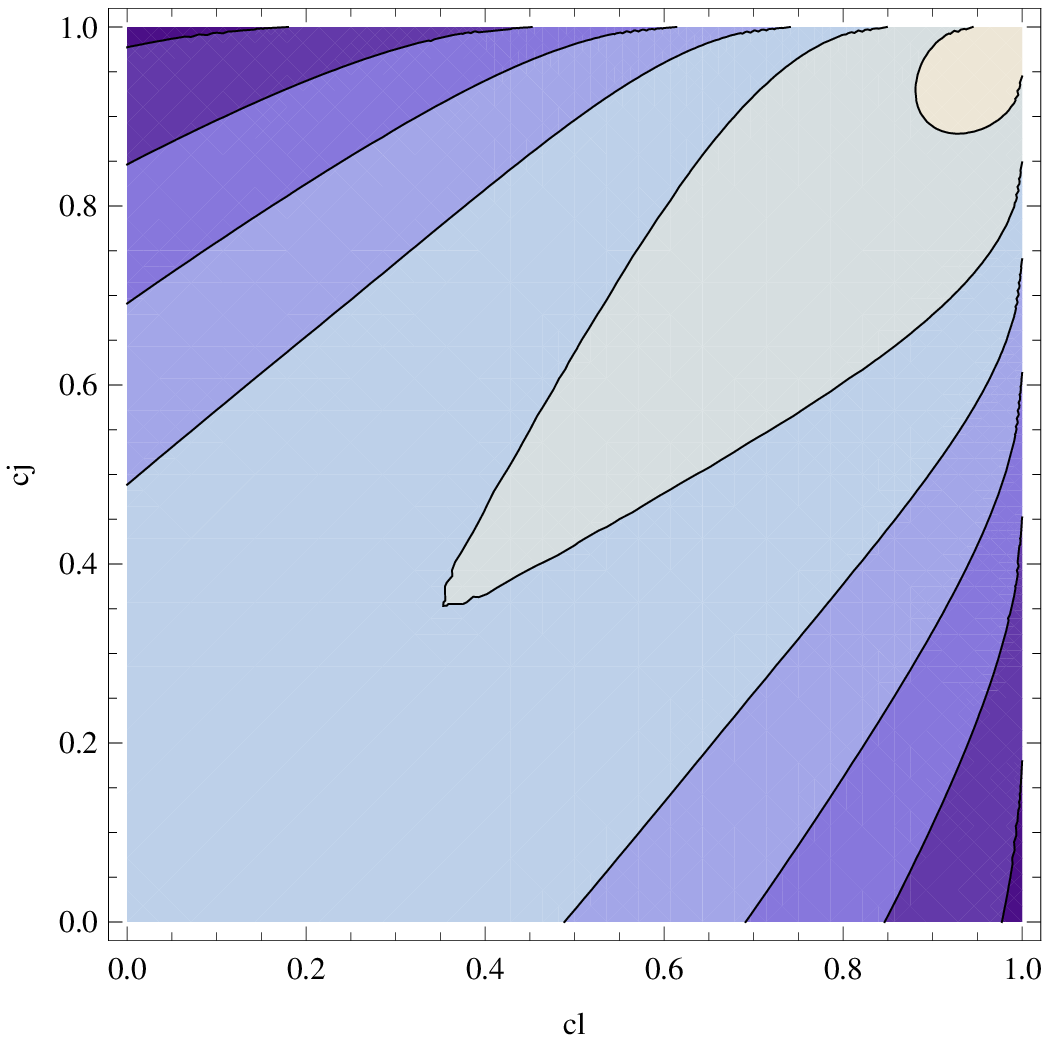,width=7cm,angle=0}
\caption{Effects of interference between longitudinal and transverse $W$ polarizations,
for $M_h = 160$ GeV and $m_{jj} > 65$ GeV (corresponding to $b = 0.42$).
Left panel: contour lines of 
 $\hat{f}_{1}(c_\ell, c_j)$;  since there is no $\varphi_\ell$ cut, there is no interference. 
Right panel: contour lines of  $\hat{f}_{1/2}(c_\ell, c_j)$ [see Eq.~(\ref{fhat12})];
a cut on $\varphi_\ell$ as in Eq.(\ref{azimuth}) with $\eta_\varphi = 1/2$ leads to interference, which lifts the
saddle point.
From the maximum at (1,1) where $\hat{f}_{1}=\hat{f}_{1/2}=1.16$, the first contour line is at 
$\hat{f}_{1}=\hat{f}_{1/2}=1.1$, and each of the other ones marks a change of 0.1;
the lighter shades represent higher values.  }
\label{fig:ccplot-int}
\end{figure}

Let us now turn to the case $\eta_\varphi = 1/2$, which implies that the last term in 
Eq.~(\ref{double-below}) does not vanish.
This term is induced by the interference between the longitudinal and transverse $W$ polarizations.
The Higgs decay width is now as in Eq.~(\ref{width-cuts}) with $\hat{f}_{1}$ replaced by 
$(1/2)\hat{f}_{1/2}(c_\ell, c_j)$, where 
\be
\hat{f}_{1/2}(c_\ell, c_j) = \hat{f}_{1}(c_\ell, c_j) 
+ \frac{8 I_1}{\pi (2 I_2 + I_0)} s_\ell c_\ell s_j c_j ~~.
\label{fhat12}
\ee
This last term contributes mostly near the saddle point of the $\hat{f}_{1}$, so that the contour lines
change their shape, as illustrated in Figure~\ref{fig:ccplot-int}.
This effect is most notable near the $WW$ threshold. 
If a standard Higgs boson with a mass not far from $2M_W$ will be discovered, then the effects 
of interference should be observed through the
measurement of the differential decay width as a function of $c_\ell$ and $c_j$ for different $\varphi_\ell$ cuts.

\smallskip

\section{Conclusions and outlook} \setcounter{equation}{0}

The semileptonic Higgs decays  $h^0\! \to WW \! \to \ell\nu jj$
and $h^0\! \to Wjj \to \ell\nu jj$, after a mass constraint on the
leptonic $W$ is imposed,
resemble the $h^0\! \to ZZ \! \to 4\ell$ golden mode in that they are
fully reconstructible. The branching fractions for these modes are
enhanced compared to the golden mode by a huge factor, ranging
between 130 above the $ZZ$ threshold and approximately
3000 near the $WW$ threshold.

We have shown that the fully
differential width for the semileptonic decays has a relatively simple dependence on the kinematic variables 
[see Eq.~(\ref{quintuple})],
and that almost all features of the decay can be exhibited analytically, even with imposition of
cuts on the relevant variables.
In the Higgs rest frame, the decay width depends on 5 kinematic variables. Treating the leptonic $W$ in the narrow
width approximation fixes the invariant mass of the $\ell\nu$ system. Two other variables may be
taken to be the dijet invariant mass, and the azimuthal angle of the charged lepton with respect to the dijet plane.
The remaining two variables are polar angles defined in the rest frames of the
dijet system, and of the $\ell\nu$ system respectively. We have shown that there are interesting correlations between these polar
angles, especially for Higgs masses away from the $WW$ threshold (see Figure 5).

The relative neglect of these semileptonic Higgs channels in the literature, and the absence of Tevatron
results for them,
can be traced to the large background contributions from Standard Model
processes with a leptonic $W$, especially diboson production and
inclusive $W + 2j$ production. To overcome this difficulty,
it is necessary to utilize all of the distinguishing features of the signal process.
The description of the fully differential Higgs decay  presented here can be
supplemented by the
fact that the Higgs peak is reconstructed in these semileptonic decays,
as well as by kinematic differences between Higgs production
and the backgrounds. In combination, this
allows one to design multi-dimensional cuts that may separate efficiently
the signal from the background.
Because the signal involves jets, it is not sufficient to study the background
at the parton level; a theoretical comparison of the differential signal and backgrounds
involving  showering effects will be presented in an upcoming publication
\cite{DobrescuLykkenWinter2010}.

At the Tevatron, we expect the semileptonic channels for Higgs decay
to provide a significant improvement in the overall sensitivity of the Higgs search,
for a large fraction of the relevant Higgs mass range. At the LHC, the semileptonic
channels should also be regarded as promising discovery channels over a broad
mass range. This should include also the semileptonic Higgs decay
$h^0\! \to ZZ \! \to \ell^+\ell^- jj$, which can be analyzed in the same way as
the semileptonic $h^0\! \to WW$, trading a higher rate for a cleaner final state.
Furthermore, these semileptonic channels can be used to supplement the fully leptonic
golden mode in the characterization of a putative Higgs signal, extending
the program developed in \cite{DeRujulaLykkenPieriniRoganSpiropulu2010}.

All our analytic expressions for the differential decay width and various cuts
have been derived here at tree level. It turns out that
the higher order effects on the shapes of the differential distributions are small, on the
order of 5\% \cite{Bredenstein:2006ha}. Thus the analytic tree level analysis outlined here
contains all of the decay information relevant to a Higgs discovery
search or an initial characterization of a Higgs signal.

\bigskip

{\it Acknowledgments.}---We would like to thank Alvaro De Rujula, Estia Eichten, Patrick Fox,
Tao Han, Robert Harr,
Maurizio Pierini, Chris Rogan, Maria Spiropulu, Zack Sullivan, 
and especially Jan Winter for stimulating discussions.
Fermilab is operated by Fermi Research Alliance, LLC, under Contract
DE-AC02-07CH11359 with the United States Department of Energy.

\end{document}